\documentclass{article}

\usepackage{amsmath}
\usepackage{amssymb}
\usepackage{amsthm}
\usepackage[english]{babel}
\usepackage{bbm}
\usepackage[width=\textwidth]{caption}
\usepackage{cases}
\usepackage{color}
\usepackage{easymat}
\usepackage{enumerate}
\usepackage{fancyhdr}
\usepackage{framed}
\usepackage{graphicx}
\usepackage[letterpaper]{geometry}
\usepackage{hyperref}
\usepackage[utf8]{inputenc}
\usepackage{listings}
\usepackage{mathrsfs}
\usepackage{natbib}

    \DeclareMathAlphabet{\mathpzc}{OT1}{pzc}{m}{it}

\theoremstyle{plain}

\theoremstyle{definition}

\theoremstyle{remark}
\newtheorem{remark}{Remark}

\newcommand{\real}{\mathbb{R}}

\newcommand{\thetitle}{A new ECDF Two-Sample Test Statistic}
\title{\thetitle}
\author{Connor Dowd\footnote{University of Chicago, \url{https://codowd.com}}}
\date{\today}

\hypersetup{colorlinks=false,pdftitle=\thetitle,bookmarksnumbered=true}

\newcommand{\iscale}{0.4}

\begin{document}

\maketitle

\begin{abstract}
Empirical cumulative distribution functions (ECDFs) have been used to test the hypothesis that two samples come from the same distribution since the seminal contribution by Kolmogorov and Smirnov. This paper describes a statistic which is usable under the same conditions as Kolmogorov-Smirnov, but provides more power than other extant tests in that vein. I demonstrate a valid (conservative) procedure for producing finite-sample p-values. I outline the close relationship between this statistic and its two main predecessors. I also provide a public R package (\href{https://cran.r-project.org/web/packages/twosamples/index.html}{CRAN: \texttt{twosamples}}\footnote{Published 2018}) implementing the testing procedure in $O(N\log(N))$ time with $O(N)$ memory. Using the package's functions, I perform several simulation studies showing the power improvements. 
\end{abstract}

\section{Introduction}
Determining whether two samples came from the same distribution is an old problem with constant relevance. Particularly when two distributions may have the same mean, but differ in other important ways, testing their similarity can be both difficult, and critical. In this paper, I characterize a new testing procedure for this situation, which builds on the literature started by Kolmogorov and Smirnov.

Consider the following situation: there are two independent samples: $A$ and $B$, of sizes $n_a$ and $n_b$. Within each sample, all observations are independently drawn from the same distribution: $a \overset{iid}{\sim} E$ and $b \overset{iid}{\sim} F$. Our null hypothesis is that the two (cumulative) distributions are the same, $H_0: E = F$. Without making any further assumptions, we would like a valid (and ideally consistent/powerful) test of this hypothesis.

Validity in the testing setting refers to a testing procedure which has the correct rejection rate when the null is true. That is to say, when we set a critical value of 5\%, the test should only reject 5\% of the time if the null hypothesis is true. Consistency refers to the ability of the test to detect small differences in the limit. For a consistent test, there is a sample size beyond which it rejects the null with high probability, when the null is false. I prove validity for this test statistic, and I will outline a proof that the test is consistent --  able to eventually detect any differences between two CDFs.\footnote{This is weaker than being consistent for any difference between two distributions. The PDF, not the CDF, uniquely identifies a distribution. By the same token, there are different PDFs which do not cause differences in the PDF. No test based solely on the ECDF will be able to detect the difference between such distributions.} But two tests which detect a difference asymptotically may still have massively difference performance. Power is how we discuss performance differences in pre-asymptotic samples. In some situations, there are already tests which have been proven to be maximally powerful, and I will compare directly to those tests. The primary benefit of this new test is greater power across a wide range of possible differences between distributions.

There are several non-standard use cases which readers may be interested in. I discuss using weighted observations, parallelizing, and comparisons to a known null distribution in Appendix A. In Appendix B I discuss code runtime and memory usage, as well as showing some real world runtime data. 

The rest of the introduction will introduce the extant testing procedures in the literature, and compare their methods in a single example simulation. That simulation consists of one sample from a standard normal, and another sample from a $N(0.5,4)$.

\subsection{Test Statistics}

\begin{figure}
    \centering
    \includegraphics[scale=\iscale]{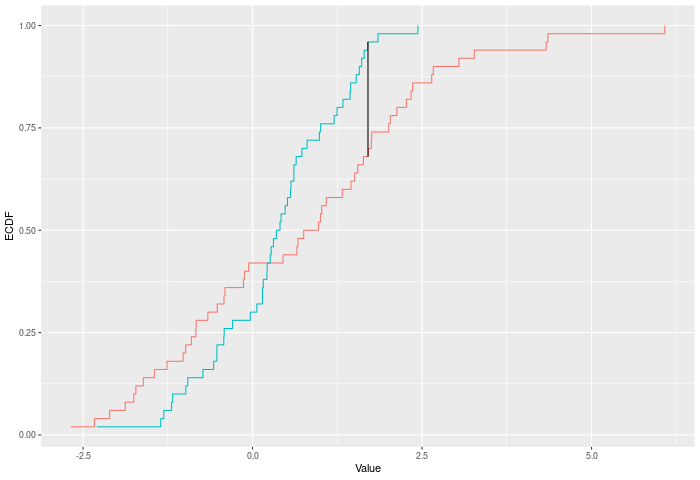}
    \caption{A demonstration of the Kolmogorov-Smirnov statistic -- the height of the black line is the KS stat. The other two lines represent the ECDFs of two independent samples.}
    \label{ks_demo}
\end{figure}

Kolmogorov and Smirnov were the first two to study this problem \cite{kolmogorov,smirnov}. Their statistic calculates the two empirical cumulative distribution functions, takes the difference, and finds the maximal absolute value of the resulting function. 
$$KS =  \underset{x \in \real}{\max} |\hat{F}(x)-\hat{E}(x)|$$
They also used innovative techniques to find the resulting asymptotic distribution, and generate p-values. Figure \ref{ks_demo} shows a black line, the height of which is the KS test statistic in that simulation. Other versions of the KS statistic are one-sided in the sense that they focused solely on $max (\hat{F}(x)-\hat{E}(x))$ or $min (\hat{F}(x)-\hat{E}(x))$ -- thus focusing their power on mean shifts in either direction. 

\begin{figure}
    \centering
    \includegraphics[scale=\iscale]{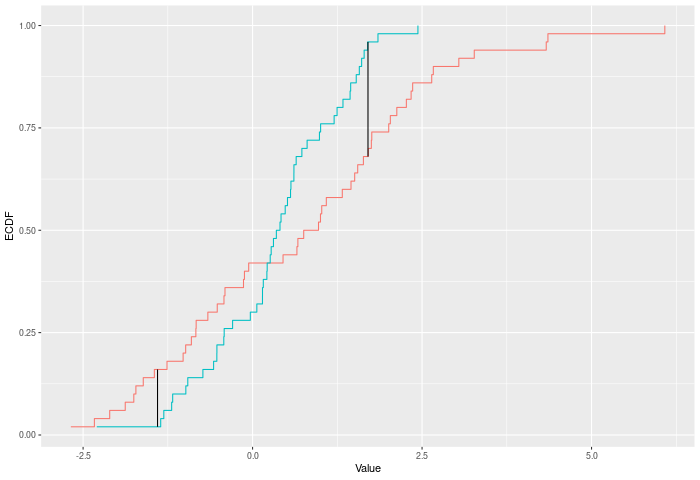}
    \caption{A demonstration of the Kuiper statistic -- the sum of the heights of the black lines is the Kuiper stat. The other two lines represent the ECDFs of two independent samples.}
    \label{kuiper_demo}
\end{figure}

Later researchers realized that by focusing solely on the maximum value, power against other hypothesis was being lost. The Kuiper test sums the max and min values of that difference.\cite{kuiper}
$$Kuiper = \underset{x\in\real}{\max} (\hat{F}(x)-\hat{E}(x)) + \underset{x\in\real}{\min} (\hat{F}(x)-\hat{E}(x))$$
This provides more power against possible variance changes -- which produce the situation in figure \ref{kuiper_demo}. The Kuiper stat would be the sum of the heights of the two black lines. 

\begin{figure}
    \centering
    \includegraphics[scale=\iscale]{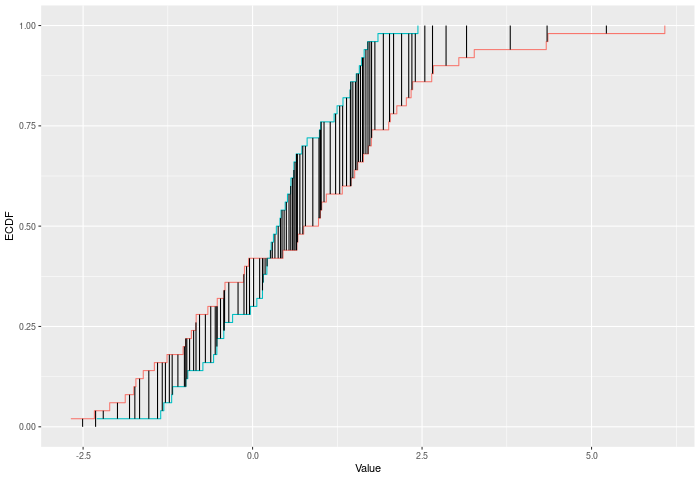}
    \caption{A demonstration of the Cramer-Von Mises statistic -- the sum of the heights of all the black lines is the CVM stat. The other two lines represent the ECDFs of two independent samples.}
    \label{cvm_demo}
\end{figure}

Cramer and Von Mises developed this further.\cite{Cramer,mises} Their test statistic takes the sum of all the observed differences between the two ECDFs. Denoting the combined sample $X$, it can be written as follows.
$$CVM = \sum_{x \in X} |\hat{F}(x)-\hat{E}(x)| $$
Figure \ref{cvm_demo} shows an example of this. The CVM test statistic would be the sum of each black line.

\begin{figure}
    \centering
    \includegraphics[scale=\iscale]{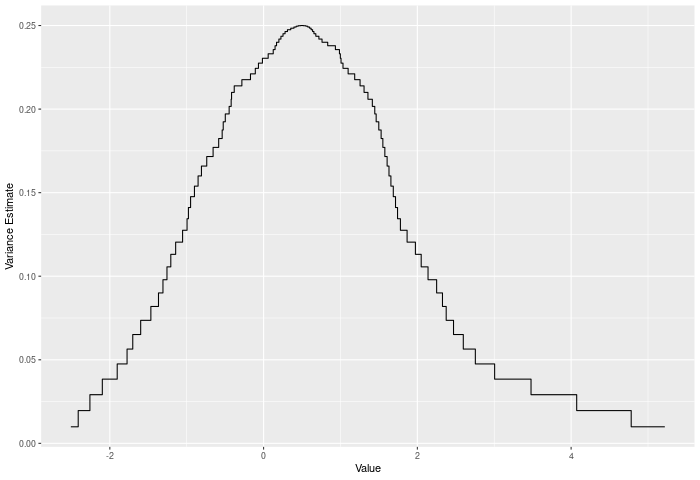}
    \caption{A plot showing how the variance of $\hat{F}(x)-\hat{E}(x)$ varies over $x$}
    \label{var_demo}
\end{figure}

\begin{figure}
    \centering
    \includegraphics[scale=\iscale]{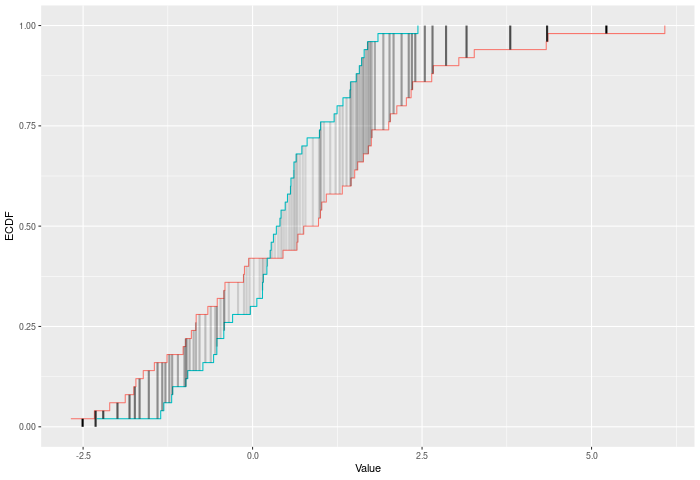}
    \caption{A demonstration of the Anderson-Darling statistic -- the weighted sum of the dark vertical lines is the AD stat. The color of those lines represents the weight each line will get. The other two lines represent the ECDFs of two independent samples.}
    \label{ad_demo}
\end{figure}

At this point, \citet{AD} noticed a central issue. Under the null, the variance of $\hat{F}(x)-\hat{E}(x)$ is not remotely stable across $x$. At any given point $x_0$, $n_a\hat{E}(x_0) \sim Binomial(n_a,E(x_0))$. That is to say, if the true CDF makes it such that $P[x < x_0] = 0.5$ for any one observation, then the ECDF, which is the fraction which are less than a given point, is distributed like a binomial with that same probability. And we know that binomial distributions have variance $np(1-p)$, which is maximized when $p=0.5$. We also know that in general, putting less weight on high variance observations, and more on low-variance observations will improve the power of a procedure. This leads naturally to the question -- how do we estimate the variance to adjust for in a two-sample version of the test? The answer is to compensate for the predicted variance of the combined sample's ECDF (denoted $\hat{D}$ below) at each point. It turns out, that under the null that the two samples come from the same distribution, this is the best estimate of the variance we can get at each point, and in the limit, it is a constant multiplier away from the true variance of $\hat{F}(x)-\hat{E}(x)$. Their estimator is:
$$AD = \sum_{x \in X} \frac{|\hat{F}(x)-\hat{E}(x)|}{\hat{D}(x)(1-\hat{D}(x))} $$
Figure \ref{var_demo} plots the predicted variances of the combined sample we've been examining. Figure \ref{ad_demo} shows a plot very similar to the CVM plot above, but where the bars are shaded by the weight they will recieve in the estimation routine. 

\begin{figure}
    \centering
    \includegraphics[scale=\iscale]{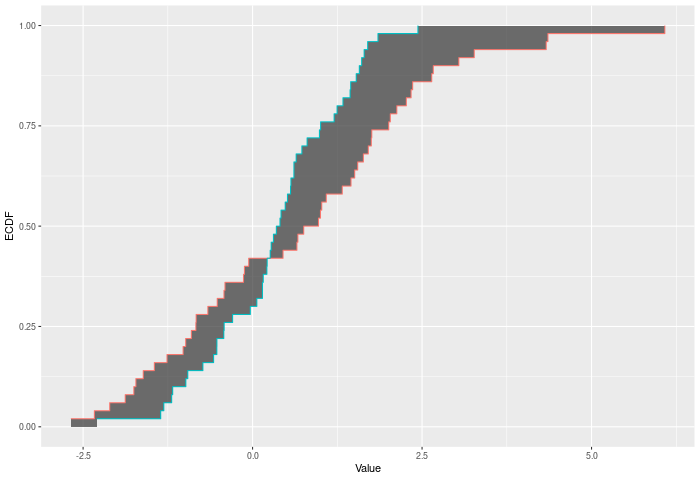}
    \caption{A demonstration of the Wasserstein statistic -- the two colored lines represent the ECDFs of two independent samples, and the Wasserstein statistic is the area between them.}
    \label{wass_demo}
\end{figure}

In a different area of statistics, Wasserstein  developed a metric closely related to optimal transport problems.\cite{wass} Broadly, the question "how little can I have to move to get from this position to that?" turns out to be related to the question of distance metrics between pdfs. The answer is not merely the sum of the distances between the ECDFs at each point, but the integral of the distance between ECDFs. Intuitively, this is a statement that the distance between observations is important and should be considered in this framework. The estimator is:
$$ wass = \int_{-\infty}^\infty |\hat{F}(x)-\hat{E}(x)| dx$$
We can see what this looks like in figure \ref{wass_demo}. 

\begin{figure}
    \centering
    \includegraphics[scale=\iscale]{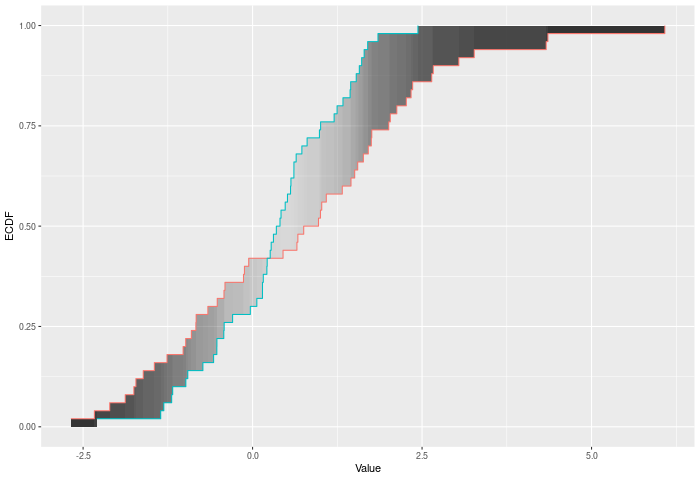}
    \caption{A demonstration of the DTS statistic -- the two colored lines represent the ECDFs of two independent samples, and the DTS statistic is the weighted integral of the their difference. The color of each region represents the weight it receives.}
    \label{dts_demo}
\end{figure}

In building up to the last two estimators, substantial power to detect differences between two distributions has been wrung out of the simple ECDF framework. However, each of the last two contributions has moved in a different direction -- each incorporating a different important bit of information. The test statistic I provide here is a synthesis of these two distinct strands in the literature. It combines the Wasserstein notion of distance as important with the Anderson-Darling realization that the variance of the estimator is changing rapidly.
$$ DTS = \int_{-\infty}^\infty \frac{|\hat{F}(x)-\hat{E}(x)|}{\hat{D}(x)(1-\hat{D}(x))}dx$$
Broadly, this serves to offer substantial power improvements in many situations over both the Wasserstein test and the AD test, as I'll show in simulations below. Figure \ref{dts_demo} shows what the test statistic looks like in an example, where the weights are being represented by the shading of the area being integrated. 

\section{Theory}
Before diving into the proper theoretical results a few notes about the applicability of the test are in order. Broadly, this test will work with any two samples that are made up of ordered data. Unlike the Chi-squared test, it cannot test categorical data. However, for data which is purely ordered, with no meaningful distances between the ranks, the Anderson-Darling test is the best bet. Nevertheless, if the implied distances between ranks are all the same (e.g. if we give the test the ranks of the observations in a sample), it can be shown that the AD test and the DTS test will give the same result. \\
As for distributional features which make this test more or less appropriate, I am not aware of any features that would break this test. Bounded domains, non-continuous PDFs, and more all lead to the same valid testing procedure.\footnote{Even though, as discussed below, some discontinuities in a PDF will be undetectable, the test will remain valid.} There are of course some assumptions that can be made allowing a more powerful testing regime. Assuming that the two samples have smooth PDFs may allow for entirely different constructions. I do not compare to such tests in simulation or the rest of this paper, as the comparison is not apples-to-apples.

\subsection{DTS test validity}
Broadly, to get from the test statistics to a full testing procedure requires finding the sampling distribution of the test statistic and then comparing our observation to that distribution. Fortunately, in our setting, this will be quite straightforward. As all the observations are independent from eachother, and under the null both distributions are the same, we get a nice exchangeability condition. Specifically, under the null, our observed samples are just as likely as any other resampling from the joint sample that results in the same sample sizes. \\
Thus any resampling is another draw from the same sampling distribution. We can generate a p-value by resampling a large number of times, and then finding the quantile in that distribution that the observed test statistic represents. Because "smaller distance between observed samples" is not indicative of anything we care about, this can be a one-sided test statistic -- i.e. we only need to compare to the right tail. \\
Further, because each of the resamples generated in this way is equally likely under the null, the rank order of our observed test statistic in the set of possible resampled test statistics is uniformly distributed over the possible rank orders. The test will only reject when there are sufficient unique test statistic values to generate a 1-in-20 event,\footnote{Or whatever level you set your critical threshold} and we observe that event. \\
Thus, under the null, the observed test statistic is drawn from a distribution of equally likely test statistics, which we can ennumerate. In order to reject the null with confidence level $\alpha$, the test statistic must be larger than $100-100\alpha$\% of the ennumerated values, which will only happen $100\alpha$\% of the time. Therefore the test is valid under the null. 

\subsection{DTS test consistency}
Proving test consistency is a bit trickier than proving test validity, so I will only try to outline such a proof. Broadly there are two questions: what is the distribution of the test statistic, and what is the distribution the test statistic is compared to. Under the null, both of these distributions are the same, which substantially eases the proof of validity. However, to show consistency, we must assess these two separately.\\
As I discussed in the introduction, this test will only be consistent for distributions which have different CDFs. This is not the same as the distributions being different, as two different PDFs can cause two different CDFs -- and be totally undetectable to testing regimes relying on the ECDF.  \\

The distribution of the test statistic is the easier of the two notions to examine. We know that ECDFs converge almost surely to the CDF of the underlying distribution, so $\hat{E}(x) \rightarrow E(x)$, and likewise $\hat{F}(x) \rightarrow F(x)$. Holding the ratio of sample sizes constant asymptotically,\footnote{This is critical for there to be a stable mixture for the joint ECDF to converge towards.} our joint ECDF also converges to the CDF of the mixture of the two distributions, $\hat{D}(x) \rightarrow D(x)$. Thus, in the limit, our test statistic $\hat{DTS} = \int |\hat{F}(x)-\hat{E}(x)|/(\hat{D}(x)(1-\hat{D}(x)))$ will converge in probability to the true distance between the distributions $\gamma = \int |F(x)-E(x)|/(D(x)(1-D(x)))$. \\This distance is strictly some positive number for all $E\neq F$. We can prove this because if $E \neq F$, there must be some range of $x$ for which $|F(x)-E(x)|/(D(x)(1-D(x)) \neq 0$. At the same time, by construction there can be no $x$ for which that same expression is negative. Thus the integral of that expression, $\gamma$, must be positive if $E \neq F$.\\
We can now state that if $E\neq F, ~~\hat{DTS} \rightarrow \gamma > 0$. However, to make claims about consistency, we need to know about the distribution to which $\hat{DTS}$ will be compared to. \\

In the resampling procedure outlined above, we draw two new samples with replacement from the combined sample, keeping sample sizes constant. Thus, with each resample we are essentially taking two independent samples from the mixture distribution characterized by $D(x)$. As $n\rightarrow \infty$, each of those samples ECDFs will converge towards the ECDF of the mixture $\hat{D}_1(x),\hat{D}_0(x) \rightarrow D(x)$. Thus for each resample $i$, the distance $DTS_i = \int |\hat{E}_i(x)-\hat{F}_i(x)|/(\hat{D}_i(x)(1-\hat{D}_i(x))) $ will converge in probability to 0. \\

The final step is to link the two statements. We know that the each resampled test statistic, $\hat{DTS}_{n,i}$ is converging in probability to 0, and we know that our observed test statistic, $\hat{DTS}_n$ is converging in probability $\gamma$, which is positive when $E \neq F$. Thus, for any critical level $\alpha > 0$ there must exist some $\bar{n}$, such that for $n > \bar{n}$, the probability that any resampled test statistic is greater than our observed test statistic is less than $\alpha/2$. At that point, when $E\neq F$, so long as we resample enough times, we will reject the null. The maximum probability of failing to reject the null for any given number of resamples is given by the probability of a binomial with $p=\alpha/2$ and $n=n_{resamples}$ exceeding $n\alpha$. For $\alpha = 0.05$, and $n_{resamples} = 100/\alpha = 2000$, this is 0.15\%.

\section{Simulation Results}
To demonstrate the strengths of each test, I've run several simulations, leveraging the speed of the \href{https://cran.r-project.org/web/packages/twosamples/index.html}{\texttt{twosamples}} package\footnote{Published 2018 on CRAN} which implements these test statistics. In the first set of simulations, I compare all the tests described above\footnote{As well as either the T-test or F-test depending on context.} as we change the amount of information observed. Two samples are generated, then each test runs on those two samples, then the whole process is repeated several thousand times to get estimates for the power of the testing procedures. 

All of the test statistics above are calculated in a standard manner, however the p-values are calculated using the resampling logic described in the theory section above, so that they should all be perfectly sized under the null. For some, such as the Kolmogorov-Smirnov test, this gives substantial power improvements relative to the usual asymptotic p-values -- which are known to be very conservative in smallish samples. It should be clear that without ensuring that all the tests had the same size control, the comparisons between tests wouldn't be fair. I do not make this adjustment for the T-test and F-test -- as they are very standardized, and I want to compare to the standard implementation. 

For each simulation exercise, I plot every test's power across the parameter that is changing. The color assigned to each test is ordered by the mean power that test had in the simulation -- so that by looking at the legend, we can see which tests performed best.

The first four simulation exercises are simple comparisons of shifted means and inflated variances -- situations which most tests should have been benchmarked on previously. Even here we see the DTS test showing performance improvements beyond its predecessors on the variance changes, while remaining competitive at detecting mean shifts. The later simulations involve both mean and variance changes, or mixtures of normals. In all of those simulations, the DTS test outperforms every other test run. At times the DTS test demonstrated a power 1.8 times as much as the best of the other advanced ECDF tests.\footnote{I exclude the Kuiper and KS tests from the `advanced' category for their abysmal performance on the first four simulations. 71\% rejection/38\% rejection = 1.87.}

\subsection{Simulations across Parameter Values}
In this section I look at two simple simulations showing the rejection rate as the difference between two sample's distributions grows. In each example, the leftmost point is a no-difference condition, so the rejection rate there represents the size of the test. As we will see, that rate consistently equals 5\% -- the nominal test size -- so all of these tests are properly calibrated. For each of these, both samples will have a sample size of 50 -- representing a total sample of 100 observations every time a test is run. 
\subsubsection{Mean Shifting}
\begin{figure}
    \centering
    \includegraphics[scale=\iscale]{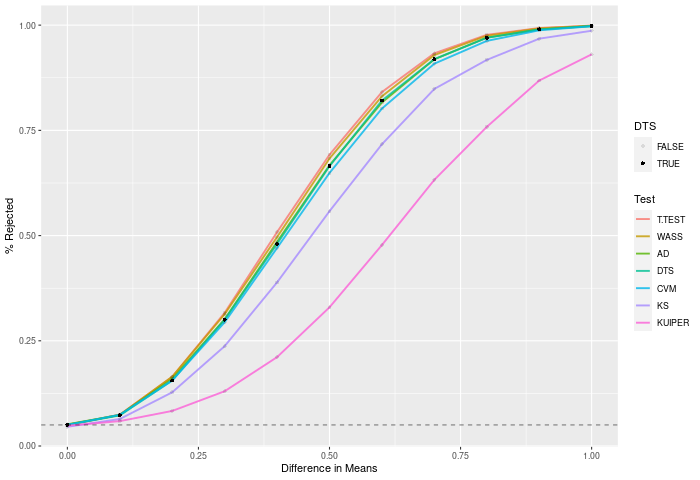}
    \caption{Rejection rates for different two-sample tests as the mean changes. $N=50$ for both samples. When the difference in means is 0, this is the test size.}
    \label{mean_shift}
\end{figure}

In this example, the first distribution is a standard normal, and the second distribution is a $N(\mu,1)$, where $\mu$ is on the x-axis. Because it is provably optimal in the circumstances, I also include a T-test. See Figure \ref{mean_shift} for results. Broadly we can see that performance for all the tests except KS and Kuiper was very comparable. 

\subsubsection{Variance Inflation}
\begin{figure}
    \centering
    \includegraphics[scale=\iscale]{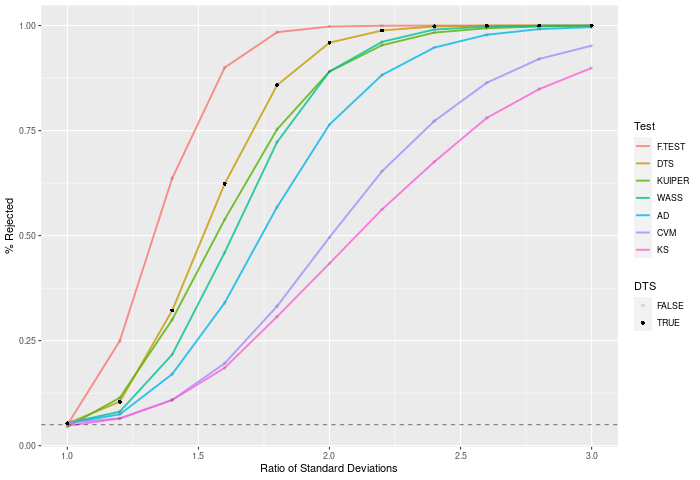}
    \caption{Rejection rates for different two-sample tests as the variance changes. $N=50$ for both samples. When the ratio of variances is 1, this is the test size.}
    \label{var_shift}
\end{figure}

In this simulation, one sample is from a standard normal, and the other is from a $N(0,\sigma^2)$, where $\sigma^2$ is on the x-axis. Because it is provably optimal in this circumstance, I also include an F-test.See Figure \ref{var_shift} for results. We see immediately that the F-test does very well. DTS follows at some distance, with Kuiper and Wasserstein a little behind it. The rest of the tests lag behind.

\subsection{Simulations Across N}
In this section I look at five simulations, as the size of both samples grow. 
\subsubsection{Mean Shift}
\begin{figure}
    \centering
    \includegraphics[scale=\iscale]{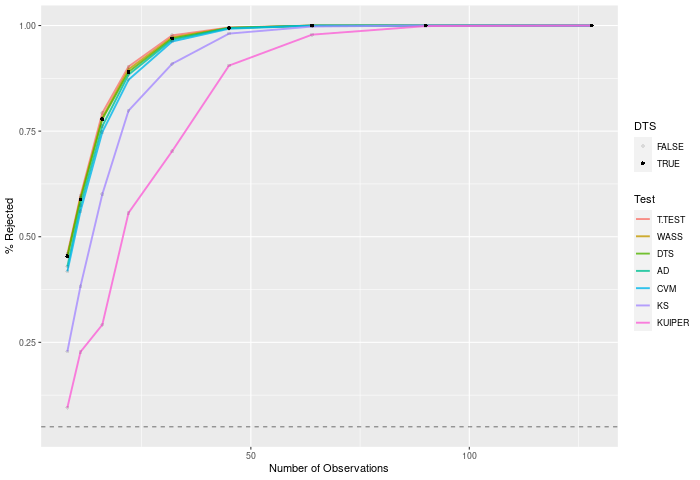}
    \caption{Rejection rates for different two-sample tests as the sample size changes. The difference in means between the samples is 1.}
    \label{n_mean}
\end{figure}

In this simulation, one sample is from a standard normal distribution, and the other is from a $N(1,1)$, i.e. the same distribution with a mean shift. Because it is provably optimal in the circumstances, I also include a T-test for comparison. 
See Figure \ref{n_mean} for results. Much like the mean simulation above, we can see that the KS and Kuiper tests lag behind, while the rest of the tests are very comparable, and indeed near the optimal power of the T-test.

\subsubsection{Variance Inflated}
\begin{figure}
    \centering
    \includegraphics[scale=\iscale]{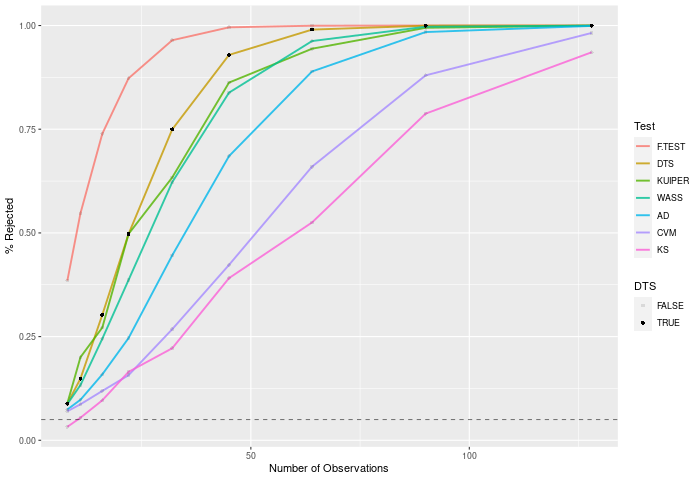}
    \caption{Rejection rates for different two-sample tests as the sample size changes. The ratio of variances between the two samples is 4.}
    \label{n_var}
\end{figure}

In this DGP, one sample is from from a standard Normal, and the other is from a $N(0,4)$, i.e. the same distribution but with larger variance. Because it is provably optimal under the circumstances, I also include an F-test for comparison. See Figure \ref{n_var} for results. Much like the variance simulation above, we can see that the F-test is far and away the best test, with DTS lagging it and followed by Kuiper and Wasserstein tests.

\subsubsection{Mean and Variance Shift}
\begin{figure}
    \centering
    \includegraphics[scale=\iscale]{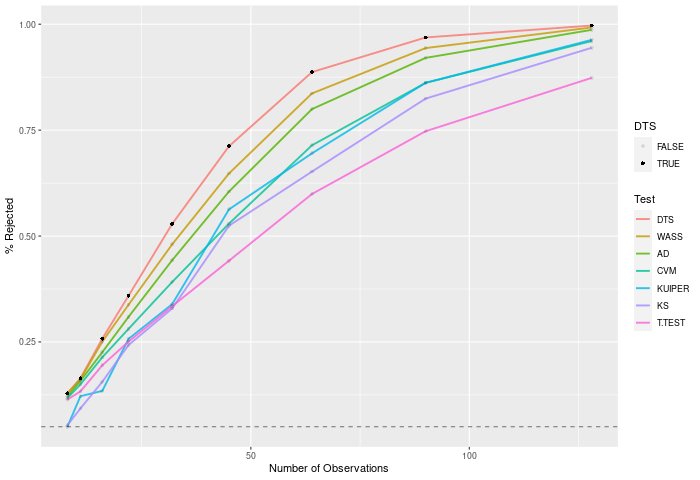}
    \caption{Rejection rates for different two-sample tests as the sample size changes. One sample is a standard normal and the other is a $N(0.5,1.5)$.}
    \label{n_mv}
\end{figure}

In this DGP, one sample is from a standard Normal, and the other is from a $N(0.5,2.25)$. I also include a $t_test$ for comparison, though strictly speaking it is not testing the same null hypothesis. See Figure \ref{n_mv} for results. In this test, we see the T-test fall away, with its performance decaying substantially relative to the competition. DTS takes a consistent lead, with Wasserstein and Anderson-Darling following. 

\subsubsection{Mixtures}
In this section I look at a much harder to detect class of distributions -- mixtures. The first sample in each dgp will be a standard normal, while the second sample in each will be a mixture of two normals. However, in order to make the problem more difficult, the mixture itself is recentered and scaled so that it has a mean of 0 and variance of 1 -- leaving only the higher moments to identify it as different from the standard normal. Because these are harder differences to detect, I've inflated the sample sizes. 

\begin{figure}
    \centering
    \includegraphics[scale=\iscale]{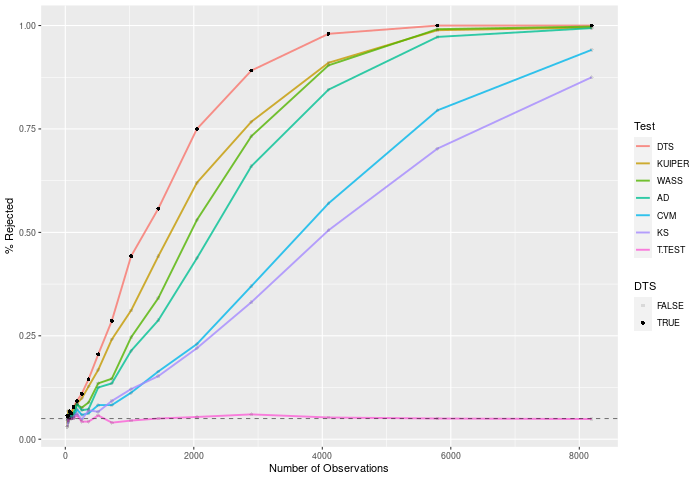}
    \caption{Rejection rates for different two-sample tests as the sample size changes. The second sample is a mixture of two normals with different means, centered so that the overall mixture has a mean of 0.}
    \label{mix_mean}
\end{figure}

The first mixture is a $N(0.8,1)$ with probability 0.2, and a $N(-0.2,1)$ with probability 0.8. See Figure \ref{mix_mean} for results. On the whole, the DTS test has the best performance, followed by the Kuiper test. As expected, the t-test detects no difference in the means of the two samples. 

\begin{figure}
    \centering
    \includegraphics[scale=\iscale]{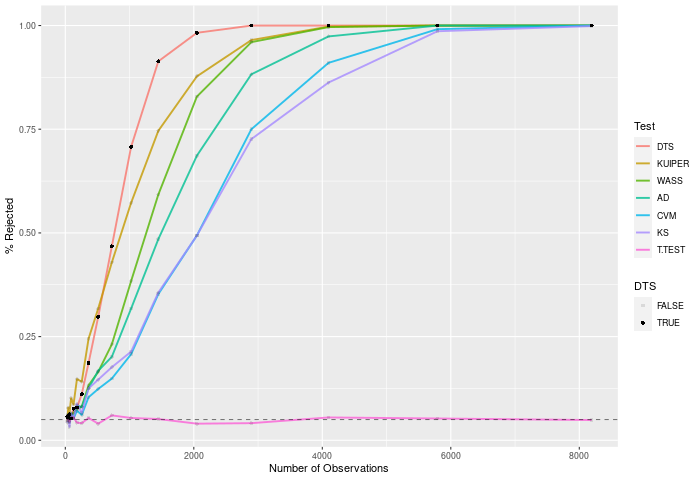}
    \caption{Rejection rates for different two-sample tests as the sample size changes. The second sample is a mixture of two normals with different variances, scaled so that the overall mixture has a variance of 1.}
    \label{mix_var}
\end{figure}

The second mixture is a $N(0,0.625)$ with probability 0.2, and a $N(0,2.5)$ with probability 0.8. See Figure \ref{mix_var} for results. Again we see the DTS test outperforming everything else, followed by the Kuiper test. By the time the DTS test had power to detect the difference 71\% of the time, the Wasserstein test was only detecting a difference 38\% of the time, and other advanced tests were even worse. Thus at that point, 1-in-3 simulations the DTS test could reject the null while other advanced tests couldn't. 

\begin{figure}
    \centering
    \includegraphics[scale=\iscale]{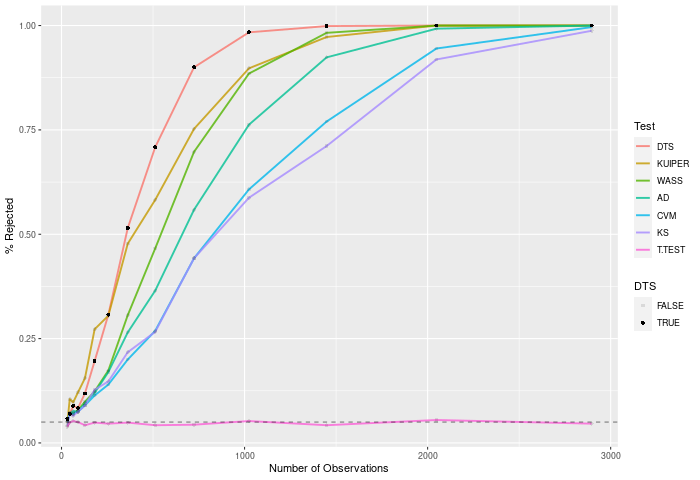}
    \caption{Rejection rates for different two-sample tests as the sample size changes. The second sample is a mixture of two normals with different means and variances, centered so that the overall mixture has a mean of 0, and scaled so that the overall mixture variance is 1.}
    \label{mix_both}
\end{figure}

The third mixture is a $N(0.8/1.7607^0.5,4/1.7607)$ with probability 0.2 and a $N(-0.2/1.7607^0.5,1/1.7607)$ with probability 0.8. See Figure \ref{mix_both} for results. Again, we see the DTS test as the most powerful, followed by the Kuiper test. Here we see that when the DTS test can reject 71\% of the time, the Wasserstein test only rejects 47\% of the time -- so that in nearly 1-in-4 samples at that sample size, the DTS test rejected while other advanced tests couldn't. 

\section{Conclusion}
The DTS test is a powerful tool for detecting differences between two samples. Particularly when we don't know how what kind of difference exists, or how much power to detect that difference we will have, the DTS test shows promise, with uniformly large power against a wide variety of distributional test statistic. As shown in simulations above, it offers power which exceeds that of other tests making comparably few assumptions about the data. Moreover, while other test statistics like the Kuiper test tended to do well in some simulations and poorly in others, the DTS test was consistently at the top, or very close to it. This suggests that while it is not uniformly more powerful than other tests, the tradeoff it manages sacrifices very little power in some situations, to gain substantial power in most situations. However, it is by no means the last word. While the DTS stat adjusts appropriately for the variance in $\hat{F}(x)-\hat{E}(x)$, it is not clear that it adjusts correctly for the variance in $\int \hat{F}(x)-\hat{E}(x)$. \\

A different perspective on the situation could be desribed as the following. The chi-squared test is proven optimal for this task when the values are sample from a distribution which takes nominal values. The Anderson-Darling test, under the same assumptions, is optimal for samples from distributions with merely ordinal values. The DTS test, under the same assumptions, offers substantial improvements towards optimality for distributions with interval values. Work towards finding the truly optimal test in this setting may require making additional technical assumptions to optimally accommodate the variance in the width of each weighted cell that the DTS stat adds. Not to mention the substantial work I've left untouched simply proving things like consistency, and power improvements over other tests. But work could also progress to the next step in the measurement framework -- looking for tests which make sense for a ratio scale. It seems likely that either taking the log of all the values before using the DTS stat, or taking the log of the widths between observations is a good step towards such a framework -- but that is a question for someone else. 

\newpage
\nocite{*}
\bibliography{dts}
\bibliographystyle{plainnat}

\appendix
\newpage
\section{Alternate Uses}
This section will address some nonstandard uses of the code in \texttt{twosamples}. Specifically, it will look at parallelization, testing against a known distribution, and weighting observations. 
\subsection{Parallelizing}
To keep things simple, \texttt{twosamples} does not internally support parallelizing the test statistics. However, the outputs from any of the \texttt{\_test} functions make it easy to run your own parallelized code. The p-values the code outputs are a count of the number of more extreme values of the test statistic observed, divided by the number of resamplings performed. If you wish to run many resamplings across cores, you could run the same code (e.g. \texttt{dts\_test(x,y)}) on each core. So long as you keep the number of resamples the same on each core, the average of your p-values will give you a new, valid p-value. If you wish to change the sample sizes, you merely need to find the appropriately weighted average of the p-values. \\

This works because when the number of resamples in each core is the same, the number of resamples automatically drops to the outside of the `correct' sum procedure. $$\frac{1}{n_{cores}}\sum_{i=1}^{n_{cores}} pval_i = \frac{1}{n_{cores}}\sum_{i=1}^{n_{cores}} \frac{extreme_i}{n_{resamples}} = \frac{1}{n_{cores}n_{resamples}} \sum_{i=1}^{n_{cores}} extreme_i $$

A small issue may arise when running the test code in parallel. When the observed test statistic is the most extreme value observed, without ties, the code defaults to a p-value of $1/(2n_{resamples})$, instead of 0. This prevents the test from rejecting the null inappropriately when either the number of observations or number of resamples is quite small. However, when performing the averaging procedure I outlined above, it would make sense to take any p-values that are $1/(2n_{resamples})$, and convert them into the more accurate 0. Then, after averaging all the p-values together, if the new p-value is 0, you should move it back up to $1/(2n_{cores}n_{resamples})$. \\

Because each core will calculate the observed test statistic independently, there is some duplication of effort. However, so long as the cores are not doing anything else important, and $n_{cores} > 2$, this can still make sense even when $n_{resamples} = 1$ on each core. 

\subsection{Weighted Observations}
Sometimes the data being used is weighted. In principle, incorporating those weights into these functions should be possible. The resampling routine can resample with appropriate weights, and then the observations contributions to the ECDF's height merely need to be adjusted to accomodate the appropriate weights. Someday, I may update the code to incorporate this ability. In the meantime however, I suggest that users desiring an ECDF test for weighted observations, find some integer $k$ such that $\min_i(k*weights_i) = 1$, and all values of $k*weights_i$ are within $0.1$ of an integer. Then, create new vectors $A$ and $B$, which contain the elements of $a$ and $b$, $k*weights_i$ times for each element. At that point you can run $dts\_test(A,B)$. This should give each observation the appropriate resampling probability and weight in the ECDF. 

\subsection{One Sample tests}
This entire paper has examined the use of \texttt{dts\_test} for the two sample problem. However, we should be able to use it to test against known distributions, like the normal. At the moment, to test whether sample A comes from a known distribution, I suggest taking $n_b = k*n_a$ draws from the known distribution to create sample B, with $k \in {10,100}$ and running \texttt{dts\_test(A,B)}. In practice, when $k=10$ this compares your sample A to a distribution which is a mixture of 91\% the known distribution and 9\% the distribution A came from.\footnote{The mixture probabilities become 99.1\% and 0.9\% when $k=100$} While there is some power lost in this procedure, relative to comparing our observed test statistic against the true null sampling distribution\footnote{This is what a true one sample version of \texttt{dts\_test} would look like}, this is a valid (and consistent) procedure which is easy to implement. In practice, particularly with large $k$, the power loss is quite minimal. 
\newpage
\section{Run Time and Memory Usage}
\begin{figure}
    \centering
    \includegraphics[scale=\iscale]{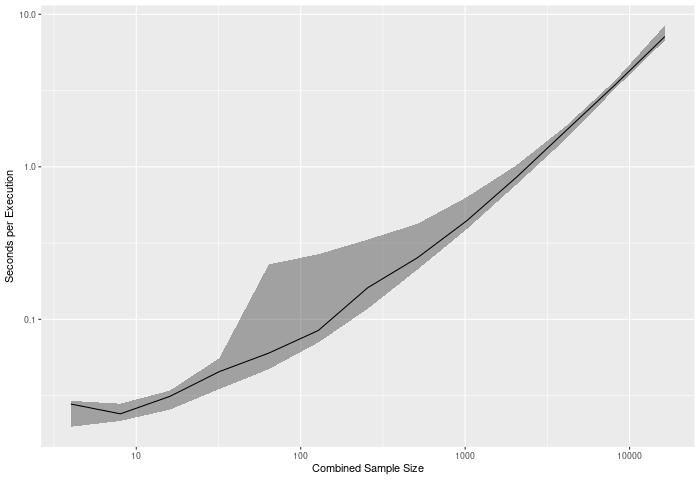}
    \caption{Execution time for \texttt{dts\_test} as $n=n_a+n_b$ grows. The black line is the mean execution time, while the ribbon represents a 95\% predictive interval for the execution time in one set of simulations.}
    \label{time_sim}
\end{figure}

The code in \texttt{twosamples} has three main components of relevance to run time. The test function consists of a loop running the test stat function $n_{resamples}$ times.\\ Each iteration of the test stat function requires one sorting of $n_a+n_b$ observations, and then another loop calculating the test statistic using that sorted data, which has $n_a+n_b$ iterations. The sort uses \texttt{std::sort}, which operates $O(n\log(n))$ on average,\footnote{Because we run this many times, the average is the main thing that matters, but \texttt{std::sort} claims an $O(n\log(n))$ worst case as well.}\footnote{Special thanks to Github user \href{https://github.com/sircosine}{sircosine} for spotting some bugs with speed.} when denoting $n=n_a+n_b$. The internals of the loop through $n$ points is not of obviously growing complexity, so this is $O(n)$ on average. Thus the test stat function operates $O(n\log(n))$ on average. Therefore the overall test function takes $O(n_{resamples}n\log(n))$ time. Of course, theory and reality rarely interact in the way we would like. Figure \ref{time_sim} shows real world execution times as sample sizes change, holding $n_{resamples}$ at 2000. 

\subsection{Memory Usage}
The memory usage of the \texttt{twosamples} package is approximately $O(N)$. The test function primarily performs a loop which operates sequentially, that loops' only side effect is to increment counters, so from iteration to iteration memory usage shouldn't change. Outside the loop the test function creates a duplicate vector of length $N$. Inside the loop, there are 4 vectors created with lengths totalling $2N$, half of which are passed on to the test stat function. That function creates 7 more vectors of length $N$, as well as about a dozen counters, for an anticipated memory use of ~$7N$. Thus the total memory usage from running the function once should be on the order of $10N$. Planned updates to this code should reduce the internal vector creation by $3N$, for a total memory usage of $O(7N)$. This is theoretical usage, not real-world measured memory, so plan accordingly. 

\end{document}